\documentstyle[12pt,epsf,lscape]{article}
\def\beq{\begin{equation}}
\def\eeq{\end{equation}}
\def\be{\begin{equation}}
\def\ee{\end{equation}}
\def\bea{\begin{eqnarray}}
\def\eea{\end{eqnarray}}

\def\lsim{\raise0.3ex\hbox{$\;<$\kern-0.75em\raise-1.1ex
\hbox{$\sim\;$}}}
\def\gsim{\raise0.3ex\hbox{$\;>$\kern-0.75em\raise-1.1ex
\hbox{$\sim\;$}}}
\begin{document}

\begin{center}
{\Large \bf Sterile neutrinos: direct mixing effects
versus induced mass matrix of 
active neutrinos}\\
\vskip 0.5cm

{Alexei Yu. Smirnov$^{a,b}$~\footnote{smirnov@ictp.it} and Renata 
Zukanovich Funchal$^{c,a}$~\footnote{zukanov@if.usp.br} 
} 
\\

\vskip 0.2cm

{\it
$^a$ ICTP, Strada Costiera 11, 34014 Trieste, Italy, \\
$^b$ Institute for Nuclear Research, Russian Academy of Sciences, Moscow,
Russia\\
$^{c}$Instituto de F\'{\i}sica, Universidade de S\~ao Paulo, C.\ P.\  
66.318, 
05315-970 S\~ao Paulo, Brazil
 }

\end{center}

\begin{abstract}
Mixing of active neutrinos with sterile ones generate ``induced''
contributions to the mass matrix of active neutrinos $\sim m_S
\sin^2\theta_{aS}$, where $m_S$ is the Majorana mass of the sterile
neutrino and $\theta_{aS}$ is the active-sterile mixing angle.  We
study possible effects of the induced matrix which can modify
substantially the implications of neutrino oscillation results.  We have
identified the regions of $m_S$ and $\sin^2\theta_{aS}$ where the
induced matrix (i) provides the dominant structures, (ii) gives the
sub-dominant effects and (iii) where its effects can be neglected.
The induced matrix can be responsible for peculiar properties of the
lepton mixing and neutrino mass spectrum, in particular, it can
generate the tri-bimaximal mixing.  We update and discuss bounds on
the induced masses from laboratory measurements, astrophysics and
cosmology.  We find that substantial impact of the induced matrix is
possible if $m_S
\sim (0.1 - 0.3)$ eV and $\sin^2\theta_{aS} \sim 10^{-3} - 10^{-2}$ 
or $m_S \geq 300$ MeV and
$\sin^2\theta_{aS} \leq 10^{-9}$.  The bounds can be relaxed 
in cosmological scenarios with low reheating  temperature, if 
sterile neutrinos decay sufficiently fast, or their masses change 
with time. 
\end{abstract}

\section{Introduction}

There are two salient properties of neutrinos related to neutrality 
which distinguish them from other known fermions: 

- the possibility to have a Majorana mass terms;  

- mix with new fermions which are singlets of the Standard Model (SM) 
symmetry  group.  

So it would seem natural to explain unusual properties of neutrinos, such
as smallness of masses and large mixing, using these two features.
In fact, the see-saw mechanism~\cite{see-saw} employs both. 

It may happen however that the see-saw mechanism is not enough to
explain the pattern of the lepton mixing, especially if quark-lepton
symmetry or unification are imposed. In this connection, we will
concentrate on the second feature - the possibility of neutrinos to
mix with singlets of the Standard Model, {\it i.e.}, sterile
neutrinos.  In general, sterile neutrinos may originate from some
other sectors of the theory, {\it e.g.}  related to supersymmetry breaking
or extra dimensions, and not coincide with the right handed components
of neutrino fields.

In all, there are two types of effects of the mixing: 

\begin{itemize}

\item
Direct effects - when new states can be produced in various processes 
and  participate in neutrino oscillations, {\it etc.};

\item Indirect effects - via the modification of the mass matrix of 
light active neutrinos.

\end{itemize}

The role and relevance of these effects is determined by masses and
mixings of sterile neutrinos.  In some ranges of parameters the 
cosmological and astrophysical consequences of mixing are more
important and the influence on the mass matrix is negligible. In other
regions {\it vice versa}: direct mixing effects are negligible - mixing
becomes ``invisible'' but the modification of the mass matrix is
substantial. 

Being light, sterile neutrinos can immediately take part in the
phenomenology of neutrino oscillations changing the interpretation of
experimental results~\cite{slight,orlando,vissani}.  Being heavy and
weakly mixed, they do not show up in oscillations and other processes,
however their ``invisible'' mixing can strongly modify the mass matrix
of active neutrinos and therefore change implications of neutrino
results for theory.  In particular, in this way, the presence of
sterile neutrinos can induce the large or maximal mixing of the active
neutrino components~\cite{mohapatra,balaji},  or, for instance, produce
deviation of the 1-2 mixing from maximal~\cite{abdel}.

In a range of masses $m_S < Q$, where $Q$ is the energy release in
processes, sterile neutrinos do not decouple since they still can be
produced. They can decouple here in the sense that their direct
dynamical effects are negligible due to the smallness of mixing.  For
bigger masses, $m_S > Q$, sterile neutrinos decouple, as it happens in
the standard see-saw mechanism. This decoupling generates an
additional contribution to the mass matrix of active neutrinos and
gives negligible deviation from universality.

From observational point of view till now there is no 
clear evidences of existence of sterile neutrinos, though  
some hints exist. Those include, the LSND result \cite{lsnd}   
and its interpretation in terms of oscillations 
in  the (3 + 1) or (3 + 2) neutrino mixing schemes;    
large scale structure formation in the Universe with  the 
warm dark matter  composed of the keV 
sterile neutrinos \cite{widrow,review}; high   
observed velocities of pulsars  and their explanation as an 
asymmetric emission of the keV sterile neutrinos~\cite{pulsar};  
the early  reionization of the universe due to the radiative 
$S$ - decay~\cite{kusenko2}.  
 
In this paper we study in details the possible effects of
sterile-active mixing on the mass matrix of active neutrinos.  We
obtain bounds on these induced masses from the direct mixing effects.
We find the impact of the induced matrix may be considerable 
if $m_S \sim (0.1 - 0.3)$~eV and $\sin^2\theta_{aS} \sim 10^{-3} -
10^{-2}$ or $m_S \geq 300$ MeV and $\sin^2\theta_{aS} \leq 10^{-9}$.

The paper is organized as follows.  In sec. 2, we determine the mass
matrix induced by mixing of active neutrinos with a sterile one.  We
study the properties of the induced mass matrix and the possibility to
explain certain features of the neutrino mass spectrum and mixing
pattern using this matrix.  We find values of mixing (as a function of
the mass of the sterile component) for which the effect of the sterile
neutrino (i) explains the dominant structures of the mass matrix; (ii)
produces the sub-dominant structures of the mass matrix; (iii) can be
neglected, being of the order or below the $1\sigma$ uncertainties of
the present measurements.  In sec. 3, we consider various bounds on
masses and mixing of the sterile neutrino and consequently, on the
induced matrix, in particular those from astrophysics and cosmology.
We then, in sec. 4, confront these bounds with regions found in sec. 2
and discuss how they can be improved in the future. We also comment on
new physics scenarios which allow to evade the bounds.  Our
conclusions are given in sec. 5.

\section{Active-sterile mixing and induced mass matrix}

\subsection{Induced mass matrix}

Let us consider three active Majorana neutrinos  
$\nu_a = (\nu_e, \nu_{\mu}, \nu_{\tau})^T$ with 
mass matrix 
\begin{equation}
{\mathbf{m_a}} =\left(\begin{array}{ccc} 
m_{ee} & m_{e\mu} & m_{e\tau}\\
m_{\mu e} & m_{\mu \mu} & m_{\mu \tau}\\
m_{\tau e } & m_{\tau \mu } & m_{\tau \tau}
\end{array}
\right)
\end{equation}
generated, {\it e.g.}, by the see-saw mechanism~\cite{see-saw}.  
We consider $m_a \leq 1$ eV -  below the present upper bound to avoid strong 
cancellations of different contributions. 

We assume that (i) the active neutrinos mix with a single (for
simplicity) sterile neutrino, $S$, via the masses  
\be m_{a S}^T
\equiv (m_{eS}, m_{\mu S}, m_{\tau S}); 
\ee 
(ii) $S$ has a Majorana
mass, $m_S$, which is much larger than the mixing masses and $m_a$:
\be m_S \gg m_{\alpha S},~ m_a.
\label{cond1}
\ee
So, in the basis $(\nu_a, S)$,  the complete mass matrix 
has the form 
\begin{equation}
\left(\begin{array}{cc}
\mathbf{m_a} & m_{a S}\\
m_{a S}^T & m_{S}
\end{array}
\right).
\label{total-m}
\end{equation}
Properties of $S$:  
masses,  mixing,  possible new symmetries, {\it etc.}, 
are determined by some new physics 
which, in general, differs  from physics responsible for 
generation of ${\bf m_a}$.

Under condition (\ref{cond1}) the block diagonalization gives for the 
light neutrinos the mass matrix  
\be
{\mathbf{m_\nu}} \approx {\mathbf{m_a}} + {\mathbf{m_{I}}},   
\ee
where 
\be
{\mathbf{m_{I}}} \equiv - \frac{1}{m_S} (m_{aS})\times (m_{aS})^T,
\ee
is the {\it induced contribution} to the 
neutrino mass matrix due to active - sterile mixing, or shortly,  
{\it  induced mass matrix}.  
For the  individual matrix element we have 
\be
(m_{\nu})_{ij} = (m_a)_{ij} - \frac{m_{iS} m_{jS}}{m_S}. 
\ee
Let us introduce the active - sterile mixing angles 
\be
\sin \theta_{jS} \approx \frac{m_{jS}}{m_S}. 
\ee
Then the induced masses  can be written as 
\be
(m_{I})_{ij} = - \sin \theta_{iS} \sin \theta_{jS} m_S.
\ee
It is this combination of parameters which determines physical effects. 
For the  flavor blind mixing we would have simply the product  
$\sin^2 \theta_{S} m_S$.

In the case of a single sterile neutrino the induced contribution is
the singular (rank-1) matrix. This feature substantially restricts 
possible effects of the induced matrix.  In the case of two (several)
sterile neutrinos, two (several) independent singular contributions to
the induced matrix appear: 
\be {\mathbf{m_{I}}} = - \sum_i
\frac{1}{m_S^{(i)}} (m_{aS}^{(i)})\times (m_{aS}^{(i)})^T.  \ee 
That opens new possibilities in the description of neutrino mass matrices.
Apparently, with three neutrinos any structure of the matrix can be
reproduced.

\subsection{Neutrino mass matrix in flavor basis}

To evaluate the impact of the active-sterile mixing, we reconstruct
the neutrino mass matrix from the data in the flavor basis in the
context of three active neutrinos.
The values of matrix elements in terms of the oscillation parameters 
are given by 
\begin{equation}
m_{\alpha \beta} = m_1 \; e^{-i 2\lambda_1} \; 
U^{*}_{\alpha 1} U^{*}_{\beta 1}+
m_2 \; U^{*}_{\alpha 2} U^{*}_{\beta 2} +
m_3 \; e^{-i 2\lambda_3} \; U^{*}_{\alpha 3} U^{*}_{\beta 3} 
\end{equation}
with $\alpha,\beta = e,\mu, \tau$, and $U_{\alpha i}$ being the elements 
of the 
PMNS matrix. The matrix elements $U_{\alpha i}$ are functions of 
the three mixing angles $\theta_{12}$, $\theta_{13}$, 
$\theta_{23}$ and 
the complex phase $\delta$ given by 
the  standard parametrization of  the mixing matrix.  

We  use the best fit 
values and  1 $\sigma$ intervals of the oscillation parameters from 
Ref.~\cite{lisi-global}:
\be
\vert \Delta m^2_{32} \vert = 2.4 \;
\left(1.00~^{+0.11}_{-0.13}\right) \times 10^{-3}~ {\rm 
eV^2}, 
\ee
\be
\Delta m^2_{21} = 7.92 \;(1.00 \pm 0.045)  \times 10^{-5} ~{\rm eV^2},
\ee
\be
\sin^2\theta_{23}= 0.44 \;(1.00~^{+0.21}_{- 0.11}),
\ee
\be
\sin^2\theta_{12}= 0.314 \;\left(1.00^{+0.09}_{- 0.075} \right),
\ee
and  for $\sin^2\theta_{13}$ we take 
\be
\sin^2\theta_{13}= 0.9\; \left(1.0~^{+3.1}_{ -0.9} \right)\times 10^{-2} 
\ee
with the non-zero best fit value. 

The reconstructed matrices for the normal, inverted mass hierarchy and
the degenerate mass spectrum are given in Table 1, where  we show the
absolute values of the matrix elements. For each case we present (i)
the matrix for the best fit values of the parameters, (ii) the
intervals which correspond to the $1\sigma$
experimental uncertainties for zero CP-phases and the intervals
when also the CP-violating phases vary in whole possible range:
$\lambda_i = 0 - \pi$, $\delta = 0 - \pi/2$ .

The following comments are in order. 

1) In the case of normal mass hierarchy we take $m_1 = 0$, so that 
$m_2 = \sqrt{\Delta m^2_{21}}$ and $m_3 = \sqrt{\Delta m^2_{31}}$.  
Notice that the 1-3 element of the matrix is much smaller than the 1-2 
element  and elements 
of the dominant block are different. This is the consequence of non-zero 
(though statistically insignificant) 1-3 mixing and shift of the 2-3 mixing 
from maximal one.  
The $1\sigma$ experimental uncertainties  lead to 
$\delta m_{ee} \sim 2.5$ meV,  $\delta m_{e\mu}  \sim  
\delta m_{e\tau} \sim \delta m_{\mu\tau} \sim 5$ meV, 
$\delta m_{\mu\mu} \sim \delta m_{\tau\tau} \sim 10$ meV. 
The effect of  CP-phases is sub-leading.  
Variations of phases $\lambda_3$, $\delta$  result in similar size of 
the intervals and both effects double the indicated uncertainties. 

Notice that there are strong  correlations  between 
elements, so that their values  can not be 
taken from the indicated intervals independently. 

2) For the inverted hierarchy we take 
$m_3 = 0$, $m_1 \approx m_2 \approx \sqrt{\vert \Delta m^2_{31} \vert}$. 
We choose $\lambda_1 = \delta = 0$ ($\lambda_3$ is irrelevant) 
for the best fit data analysis. 
The $1\sigma$ experimental uncertainties lead to 
$\delta m \sim (8 - 10)$ meV for all matrix elements but 
$\delta m_{\mu\tau} \sim 4$ meV. The effect of phase variations 
is much stronger. Now the phases affect the dominant block 
and therefore $\delta m \sim m$.  

3) Degenerate spectrum. For illustration we take 
$m_1 \approx m_2 \approx m_3 = m_0 = 0.2$ eV. 
The experimental errors produce very small effect: 
$\delta m \sim 1 $ meV. In contrast, effects of phase variations are very 
strong, $\delta m \sim m_0$. Here also variations (values) of different 
elements are strongly correlated.



\subsection{Induced matrix and the dominant structures}

The mass and mixing patterns in quark and lepton sectors are strongly
different. The difference can (at least partially) originate from the
active-sterile mixing which is absent in the quark sector.  The
shortcoming of this proposal is the coincidence problem: two different
contributions to the mass matrix, active, $\mathbf{m_a}$, and induced,
$\mathbf{m_I}$, are of the same order or within 1-2 orders of
magnitude in spite of the fact that they have different, and at the
first sight, unrelated origins.  The only argument in favor is that
this will be not the only case - we meet the coincidence problem in
other areas too. Another possibility is that $\mathbf{m_I} \gg 
\mathbf{m_a}$.  

Let us consider first the case of one sterile neutrino.  Due to the
singular (rank 1) character of the induced matrix it  can not
reproduce the dominant structures of the neutrino mass matrix in the
case of the degenerate mass spectrum and inverted mass hierarchy.  In
the former case ${\rm Det}\,{\bf m} \approx m_0^3$, where $m_0$ is the
scale of neutrino mass.  In the latter - there are two dominant
eigenvalues and the determinant of the 1-2 submatrix is non-zero.


Essentially this means that the matrix induced by one sterile neutrino
can be the origin of the dominant block {\it only} in the case of
normal mass hierarchy.  This can explain the large (maximal) 2-3
mixing.  Suppose \be m_{\mu S} = m_{\tau S} = m_0, ~~~m_{eS} \ll m_0,
\label{domblock}
\ee
then
\begin{equation}
{\mathbf{m_{I} }} \approx 
\frac{m_0^2}{m_S}
\left(\begin{array}{ccc} 
0 & 0 & 0\\
0 & 1 & 1\\
0 & 1 & 1
\end{array}
\right). 
\label{matrdom}
\end{equation}
From (\ref{matrdom}) we obtain $2m_0^2/m_S = \sqrt{\Delta m_{\rm
    atm}^2}$ or \be m_S \sin^2 \theta_S = 0.5 \, \sqrt{\Delta m_{\rm
    atm}^2} \approx 25~{\rm meV}.
\label{atm}
\ee 

Parameters of the 1-2 block (mixing and mass)  
should be given by the original active mass  matrix.  
For the best fit values of the oscillation parameters 
(here we take $\sin^2\theta_{13} = 0$) and the induced 
matrix (\ref{matrdom}) the active neutrino mass matrix should be 
of the form 
\begin{equation}
{\bf{m_a}} = U_{23}^{\rm m} \, U_{12}^{\rm sol} \, {\mathbf{m_2}}\,  
U_{12}^{{\rm sol} T} \, U_{23}^{{\rm m}T} = 
m_2 \, 
\left(\begin{array}{ccc}
s^2 & \frac{sc}{\sqrt{2}} & - \frac{sc}{\sqrt{2}}\\
... & \frac{c^2}{2}  & -\frac{c^2}{2}\\
... & ... & \frac{c^2}{2} 
\end{array}
\right), 
\label{matrsub1}
\end{equation}
where ${\mathbf{m_2} = diag}(0, m_2, 0)$, $s \equiv \sin \theta_{12}$,
$c \equiv \cos \theta_{12}$, $U_{12}^{\rm sol}$ is the 1-2 rotation matrix on
the solar mixing angle, $U_{23}^{\rm m}$ is the matrix of maximal 2-3  
mixing.  
We assumed that $m_1
\approx 0$.  Notice that all elements of this matrix are nearly equal
being  in the range 0.31 - 0.35.

A non-zero $m_1$ would be equivalent of adding to (\ref{matrsub1}) the
matrix $m_1 {\mathbf I}$ proportional to the unit matrix and
substituting $m_2 \rightarrow (m_2 - m_1)$, if there is no
CP-violating phases.  That does not change our conclusion provided
that $m_1 \ll m_2$.

The induced  matrix can be chosen in such a way that 
the active one has hierarchical structure with small mixings 
similar to the charged fermion mass matrices. (Partly this 
case has been studied in \cite{balaji}). 
Consider non-universal active-sterile coupling of the 
type 
\be 
m_{\alpha S} = \sqrt{(m_3 - m_1) m_S} ~(\alpha, a, b),~~~ \alpha \ll a 
\sim b. 
\ee
Then the required mass matrix of active neutrinos can  be written as 
\begin{equation}
{\mathbf{m_{a}}} = 
m_1 {\bf I} + (m_3 - m_1)
\left(\begin{array}{ccc}
\epsilon s^2 - \alpha^2 & \epsilon \frac{sc}{\sqrt{2}} - \alpha a   & - 
\epsilon \frac{sc}{\sqrt{2}} - \alpha b\\
... &1 - a^2 + \epsilon\frac{c^2}{2}& 1 - ab - \epsilon \frac{c^2}{2} \\
... & ... & 1 - b^2 + \epsilon \frac{c^2}{2}
\end{array}
\right),
\label{matrsub}
\end{equation}
where $\epsilon \equiv  m_2/(m_3 - m_1)$. 
For $a = 0.95$ and $b = 1.50$ we obtain hierarchical 2-3 block 
with $(m_a)_{22} : (m_a)_{23} : (m_a)_{33} \approx 0.1 : 0.4 : 1.2$. 
Other  elements are of the order 0.02 - 0.04 if 
$\alpha \sim 0.01$. Certain hierarchy among those elements can be obtained 
for non-zero 1-3 mixing and some deviation of the 2-3 mixing from 
maximal. The overall spread of the values of elements 
by 2 orders of magnitude can be easily obtained. 

Another scenario (essentially considered in \cite{balaji}) is that $b
\gg a$, so that the induced matrix corrects the 33
element only. Taking, {\it e.g.}, $m_{33} = \sin^2 \theta_{\tau S} \, m_S
\sim 300$ meV and the final matrix $\mathbf{m_{\nu}}$ as in the Table
1, we obtain that the original active mass matrix should be like that
for normal hierarchy (Table 1) but with $m_{33} \sim 300$ meV,
that is, strongly hierarchical.

\subsection{Induced matrix and mass hierarchy}

An interesting possibility is that the induced matrix can switch the
mass hierarchy from normal to inverted and {\it vice versa}. Indeed,
\be
{\mathbf{m_{\nu}}^{\rm inv}} \sim \sqrt{2} \,  {\mathbf{m_{\nu}^{\rm norm}}} 
- \sqrt{\frac{\vert \Delta m^2_{32}\vert}{2}} \mathbf{D}, 
\ee
where the induced term, $\mathbf{D}$, is close to the ``democratic'' matrix 
with all elements being nearly  1. 

\subsection{Induced matrix and tri-bimaximal mixing}

The mass matrix which generates the tri-bimaximal mixing 
in the normal mass hierarchy case can be presented as 
\begin{equation}
{\mathbf{m}} =
\frac{\sqrt{\vert\Delta m_{32}^2}\vert}{2} 
\left(\begin{array}{ccc} 
0 & 0 & 0\\
0 & 1 & - 1\\
0 & - 1 & 1
\end{array}
\right) + 
\frac{\sqrt{\Delta m_{21}^2}}{3}
\left(\begin{array}{ccc} 
1 & 1 & 1\\
1 & 1 & 1\\
1 & 1 & 1
\end{array}
\right). 
\label{tribi}
\end{equation}
It is the sum of two singular matrices.
The sub-dominant (second)  matrix can 
be induced by universal mixing with the sterile 
component  
\be
m_{\alpha S} =  m_0 (1, 1, 1). 
\ee
Then according to (\ref{tribi}) 
\be
m_S \sin^2 \theta_S = \frac{\sqrt{\Delta m_{21}^2}}{3} 
\approx 3~{\rm meV}.
\label{sol}
\ee 

In fact, both matrices in (\ref{tribi}) can be induced by the 
active-sterile mixing, if the second sterile neutrino 
is introduced with mixing elements
\be
m_{\alpha S}' =  m_0 (0, 1, -1) . 
\ee
In this case the original active neutrino masses should be very small,  
{\it e.g.},  of the order of the Planck mass suppressed scale: 
$\sim v_{\rm EW}^2/M_{\rm Pl}$.  

\subsection{Induced matrix and QLC}

If not accidental, the quark-lepton complementarity (QLC)
relation~\cite{qlc} can imply that (i) there is some structure in the
lepton sector which generates the bi-maximal mixing $U_{\rm bm} \equiv
U_{23}^{\rm m}\, U_{12}^{\rm m}$; and  (ii) there is the quark-lepton symmetry 
which
``propagates'' the CKM-type rotations to the lepton sector.

Let us consider a possibility that the induced matrix is responsible
for the bi-maximal mixing: 
\be {\bf m_I} = {\bf m_{bm}} = U_{\rm bm} {\bf
  m^{diag}} U_{\rm bm}^T, \ee 
whereas the charged lepton mass matrix
produces the CKM-type rotation.  The original active neutrino matrix
should then give very small contribution.

Clearly this scenario can not be realized with only one sterile 
neutrino: Taking 
${\bf m^{diag}} =  {\bf diag}(0, 0, m_3)$ we find that 
${\bf m_I} = U_{23}^{\rm m}\, {\bf m_3} \, U_{23}^{{\rm m}T}$ 
with a single maximal mixing. 
With two sterile neutrinos the matrix  ${\bf m_{bm}}$ 
which generates the bi-maximal mixing can be reproduced 
precisely. As an example one can take 
$m_{\alpha S} =  m_0 (a, b, -b)$ and 
$m_{\alpha S}' =  m_0 (0, x, y)$, where 
$x = \sqrt{d + 0.5a^2  - b^2}$,  $y = (d - 0.5a^2  + b^2)/x$, 
and $a, b, d$ are free parameters.

\subsection{Small and negligible induced contribution}

The induced matrix becomes irrelevant if
\be
\frac{m_{iS}m_{jS}}{m_S} \ll (m_a)_{ij}. 
\label{cond}
\ee

Let us find conditions under which the effects of  sterile neutrino
are below the present $1 \sigma$ spread of the matrix elements.
Uncertainties depend on the type of mass spectra and are different for
different elements.

For the normal mass hierarchy the effects are below $1 \sigma$ if 
\be
\sin^2 \theta_{eS} \, m_S < 2~{\rm meV}, ~~
\sin^2 \theta_{\mu S} \, m_S,~~ \sin^2 \theta_{\tau S}\, m_S  < 5~{\rm meV}, 
\label{normmix}
\ee
For $\mu$ and $\tau$  we have taken the uncertainty of 
$m_{\mu \tau}$ which is  the smallest one. 

For the inverted hierarchy the induced contributions are 
below $1\sigma$ uncertainties if  
\be
\sin^2 \theta_{eS} \, m_S < 8~{\rm meV}, ~~
\sin^2 \theta_{\mu S} \, m_S < 4 ~{\rm meV}. 
\label{invmmix}
\ee
Larger effects of sterile neutrinos,  
$\sin^2 \theta_{S} m_S < 20$~{\rm meV},  
can be mimicked by the phase variations. 
  
In the case of degenerate spectrum $1\sigma$ 
experimental uncertainties restrict 
\be
\sin^2 \theta_{eS} \, m_S < 1~{\rm meV}. 
\ee
The phase change produces the same effect as sterile neutrinos with 
$\sin^2 \theta_{\alpha S} \, m_S \sim 200$ meV. 

Sterile neutrinos can be responsible for fine structures of the mass
matrix.  The smallest (observable) structure is related to the solar
mass split.  In the case of inverted hierarchy this would correspond
to the contribution $\sin^2 \theta_{eS}\, m_S \sim 1.5$~meV, and for
the degenerate mass spectrum we obtain the smallest quantity: $\sin^2
\theta_{eS}\, m_S \sim 0.4$~meV.

So, we can identify three benchmarks: 

1) For sterile neutrinos with  mixings and masses 
smaller than 
\be
\sin^2 \theta_{\alpha S}\, m_S = 1~{\rm meV} ~~
\label{small}
\ee
the effects are below the present $1\sigma$ experimental 
uncertainties for the hierarchical spectra. 
Still  these neutrinos can influence the sub-leading structures in the 
case of the degenerate spectrum.  

2) Sterile neutrinos with 
\be
\sin^2 \theta_{\alpha S}\, m_S = 3~{\rm meV} ~~
\label{sublead}
\ee
can generate the sub-leading structures in the 
case of normal mass hierarchy.

3) Sterile neutrinos with
\be
\sin^2 \theta_{\alpha S}\, m_S = (20 - 30)~{\rm meV} ~~
\label{leadA}
\ee
can generate dominant  structures in the
case of normal and inverted hierarchies. 
For larger masses and mixings they can lead to  dominant structures of 
the degenerate spectrum.  

In Figs.~\ref{fig1}, \ref{fig2} and \ref{fig3} we show the lines of
constant induced mass $\sin^2 \theta_{\alpha S}\, m_S =  {\rm const.}$, 
in the plane $\sin^2 \theta_{\alpha S}$ and $m_S$ which correspond to
the values in (\ref{small}, \ref{sublead}, \ref{leadA}).

\section{Bounds on the active-sterile mixing}

In this section we describe the direct mixing effects of $S$ and    
bounds on its masses and mixings.  

\subsection{Production, thermalization, decay}

The most stringent bounds follow from astrophysics and cosmology. 
Sterile neutrinos can be produced in the Early Universe non-thermally
through their mixing with active neutrinos, affecting primordial
nucleosynthesis, cosmic microwave background radiation as well as the
growth of cosmological structures.

We assume that no primordial density of sterile neutrinos existed, 
and all sterile neutrinos where produced in the Early Universe 
due to mixing with active neutrinos and oscillations~\cite{widrow,review}. 

The ``thermalization'' lines and the ``decay'' lines in the $m_S - \sin^2
\theta_S$ plane (Figs. \ref{fig1}-\ref{fig3}) allow to understand
various bounds.

1) The thermalization lines give the lower bounds of the $m_S -
\sin^2 \theta_S$ region where the sterile neutrinos are thermalized
before the primordial nucleosynthesis.  According to
Ref.~\cite{enqvist,shi} the lines are slightly different for mixing with
electron and non-electron neutrinos: 
\be m_S \, \sin^2 \theta_S = 0.6 \; \rm meV, \quad {\rm for} \; 
\nu_e,  
\ee 
\be m_S \, \sin^2
\theta_S = 0.4 \; \rm meV, \quad {\rm for} \; \nu_\mu,
\nu_\tau.  
\ee 
Notice that the thermalization lines have the same
functional dependence as the isolines of  induced mass.
Furthermore, the lines are below the benchmarks obtained in
(\ref{small}, \ref{sublead}, \ref{leadA}).  That is, sterile neutrinos
with parameters which give significant induced contribution were
thermalized in the Early Universe.

2) We confront  the life time of neutrinos with the time of recombination, 
$\tau_{\rm rec}  \approx 10^{12}$ s, and 
the age of the Universe,  $\tau_{\rm U} = 4 \times 10^{17}$ s.   
The decay rate of $S$ strongly depends on $m_S$. For 
small $m_S$ the main channel is $S \to 3 \nu$ and then 
the following channels open: 
$S \to \nu + l + \bar l$, $S \to \nu + q + \bar q$, {\it  etc.}.  
The decay rate can be written as 
\be 
\frac{1}{\tau_S} \approx \kappa(m_S)  \Gamma_\mu  \left 
    (\frac{m_S}{m_\mu}\right)^5 \sin^2 \theta_S, 
\ee
where $m_\mu$ and $\Gamma_\mu$ are the muon mass and decay rate,  
correspondingly, and $\kappa(m_S)$ is the number of decay channels 
for a given value of  $m_S$.  So, for the non-relativistic sterile 
neutrinos, the lines of constant decay rate are given by 
\be
\kappa(m_S)~ m_S^5 \, \sin^2 \theta_S = {\rm const}. 
\ee

In Figs.~\ref{fig1}, \ref{fig2} and \ref{fig3} we show two decay isolines 
corresponding to $\tau_S = \tau_{\rm rec}$  and $\tau_S = \tau_U$.   

\subsection{LSS formation bound}  

If $\tau_S > \tau_U$, the sterile neutrinos  contribute to the dark
matter in the Universe.  Analysis of the Large Scale Structure (LSS) of
the Universe gives the bound on the total energy density in sterile
neutrinos, $\rho_{S}$, as function of its mass:
\be
\omega_S \equiv \frac{\rho_{S}}{\rho_{\rm cr}} h^2 \leq \omega_S (m_S). 
\label{omega-s}
\ee Here $\rho_{\rm cr}$ is the critical energy density and $h$ is the
Hubble constant\footnote{In this paper in numerical estimations we use
h=0.7.}.  We use for $\omega_S (m_S)$ the results of the analysis in
\cite{dms} and \cite{slosar}.

For $m_S < 100$ eV, $S$ compose the hot dark matter component and the
bound on $\omega_S$ is stronger.  According to \cite{dms} for
$m_S < 30$ eV one has $\omega_S < 0.005$ at $95 \%$ CL.  The bound
weakens with the increase of $m_S$, 
as neutrinos become colder: $\omega_S < 0.02$ at $m_S = 100$
eV, and $\omega_S < 0.12$ at $m_S = 300$ eV.  In the interval
$m_S=(0.25 - 30)$ eV the 95\% CL bound from \cite{slosar} can be 
parametrized as 
\be 
\omega_S \leq 0.001 \displaystyle \left[\left(\ln\left(\frac{m_S}{\rm
        eV}\right)-1.7\right)^2 +2.5\right].  
\label{lim1}
\ee 
For $m_S=(30 - 300)$ eV~\footnote{A more accurate 
parametrization in this range is:  
$$
\omega_S \leq 6 \cdot 10^{-3} + 4.1 \cdot
10^{-6} \left(m_S/{\rm eV}\right) + 2.5 \cdot 10^{-7}
\left(m_S/{\rm eV}\right)^2 + 1.7 \cdot 10^{-8}
\left(m_S/{\rm eV}\right)^3 - 4.5 \cdot 10^{-11}
\left(m_S/{\rm eV}\right)^4$$.}:
\be 
\omega_S \leq 0.01 \displaystyle \left[\left(\ln\left(\frac{m_S}{\rm
        eV}\right)-2.55\right)^{2.2}-1.0\right], 
\label{lim2}
\ee
and $\omega_S< 0.12$ for $m_S>$ 300 eV.

Calculations of the  energy density $\omega_S$  have been 
updated recently~\cite{KA} with inclusion of a number of additional effects, 
in particular, effects of the quark-hadron transition, modification of the 
finite temperature effective mass of the active neutrinos, {\it etc.}. 
For the temperature of the quarks-hadron transition $T_{QCD} = 170$ MeV, 
the relation between $\omega_S$, $m_S$ and $\theta_S$ can be parametrized 
as~\cite{KA}  
\be
m_S = 1.45~{\rm  keV} \left(\frac{10^{-8}}{\sin^2 \theta_S}\right)^{0.615} 
\left(\frac{\omega_S(m_S)}{0.13}\right)^{0.5},  
\label{a-limit}
\ee
or
\be
\sin^2 \theta_S = 1.83 \cdot 10^{-8} 
\left(\frac{\omega_S(m_S)}{0.13}\right)^{0.813}
\left(\frac{1~{\rm  keV}}{m_S}\right)^{1.626}. 
\label{a-limit2}
\ee
Notice that for $m_S < 3$ keV it deviates from the dependence previously found 
in \cite{fuller}. 

Plugging the limits (\ref{lim1}) and (\ref{lim2}) in   
relation  (\ref{a-limit2}) we find the upper bound on 
$\sin^2 \theta_S$ as a function of $m_S$  
(see region LSS in Figs. \ref{fig1}-\ref{fig3}).  
The bound is absent for $m_S < 0.25$ eV~\cite{seljak}. 
It does not depend on the flavor of  active neutrino to 
which sterile neutrino mixes. 

The bound is valid for the region of parameters below the isoline 
$\tau_S = \tau_U$ which corresponds to $m_S < 0.5$ MeV and 
$\sin^2 \theta_S > 7 \cdot 10^{-13}$. However for $m_S > 5$ keV the
stronger bound follows from  the diffuse background radiation.

It has been shown in a recent analysis~\cite{uros} that sterile
neutrinos with mass $m_S < 14$ keV are excluded at 95\% CL as the dark
matter particles responsible for the LSS formation.  This implies
somehow stronger bound on $\omega_S$ than the one we use in our
estimations and consequently stronger bound on the $S$ parameters.  It
is argued in that the bound can be relaxed if new particle decays in
the epoch between the decoupling of $S$ and BBN increase the entropy,
thus diluting concentration of $S$ and reducing their relative
temperature~\cite{asaka}.  This may be the case for the model proposed
in \cite{Asaka:2005pn,Asaka:2005an} where two heavy RH have masses
above 1 GeV.  (Explanation of LSS by $S$ would probably require that
$S$ are produced at some high energy scales by processes that are not
related to active-sterile neutrino oscillations~\cite{asaka}.)

\subsection{Limit from cosmic X-ray radiation}

Due to mixing via the loop diagrams sterile neutrinos decay into an
active neutrino and a photon: $S \rightarrow \nu_a \gamma$ with
$E_\gamma \approx m_S/2$ and at the rate $\Gamma_{\gamma} \sim \alpha
\Gamma_{3\nu}$.  Therefore one expects to detect the photon emission
line when looking at big concentrations of dark matter such as galaxy
clusters.  

The analyzes of the X-ray emission from the Virgo cluster
~\cite{tucker,ferrara} give limits on the decay and therefore on 
the mixing of  sterile neutrino: 
$\sin^2\theta_S < 2.6 \times 10^{-6} \left({m_S}/{\rm keV} \right)^{-4}$. 
(This parametrization is valid for $m_S = (1 - 10)$ keV.) 
It is argued in \cite{boyarsky} that the bound from Virgo  
is about 1 - 2 orders of magnitude weaker  especially for 
low  masses, $m_S < (5 - 8)$ MeV. However for $m_S = 10$ keV,  
relevant for this analysis  (where X-ray bounds start to  
dominate over the LSS one), the bounds are comparable: 
$\sin^2\theta_S < 2.6 \times 10^{-10}$ \cite{tucker} and 
$\sin^2\theta_S < 5.0 \times 10^{-10}$  \cite{boyarsky}.  
 At the same time stronger limit  has been obtained from 
analysis of Coma cluster  \cite{boyarsky}:  
\begin{equation}
\sin^2\theta_S < 2 \times 10^{-5} \left(\frac{m_S}{\rm keV} 
\right)^{-5}.
\label{xrad}
\end{equation}

The $\gamma$-flux from all possible sources could accumulate over the
history of the Universe and be seen as a Diffuse Extragalactic
Background RAdiation (DEBRA). Apparently the flux from the radiative
$S$ decay should be smaller than the observed flux. 
In the range $m_S = (1 - 100)$ keV  the DEBRA limit can be 
parametrized as~\cite{shaposhnikov}
\begin{equation}
\sin^2\theta_S < 3.1 \times 10^{-5} \left(
 \frac{m_S}{\rm keV} \right)^{-5}. 
\label{eq:debra}
\end{equation}
This bound based on the data collected from the whole sky 
 is weaker than the one given in (\ref{xrad}). However, it 
does not depend on assumptions concerning clustering and 
therefore is considered to be more robust~\cite{shaposhnikov}. 
These limits are valid below  the recombination line $\tau_S = \tau_{\rm rec}$.

The X-ray exclusion region shown in Figs.~\ref{fig1}-\ref{fig3} has
been obtained from the analysis of Coma cluster data as well as data
on diffuse X-ray background (DEBRA) from HEAO-1 and XMM-Newton
missions in~\cite{shaposhnikov}.  We have also included
 the very recent bound from the diffuse X-ray spectrum of the Andromeda 
galaxy~\cite{watson}. This is the most stringent limit in the 
range $m_S=(1-24)$ keV.
For small masses, $m_S \lsim 1$ keV,
we use the limit from~\cite{hansen}.

Note in Figs.~\ref{fig1}-\ref{fig3} that the X-ray data reduces
  the parameter space allowed for Warm Dark Matter to a very small
  region: $m_s=(1.7-3.5)$ keV for $\sin^2\theta_S \sim 10^{-9} -
  10^{-8}$. Also most of the parameter region that can explain the 
origin of pulsar velocities is ruled out, remaining only a corner 
around $m_S \sim 2-4$ keV.

It was proposed in Ref.~\cite{tucker} to observe clusters of galaxies
with Chandra and XMM-Newton observatories, in their high sensitivity
range for X-ray photon detection of (1 - 10) keV.  That will allow one
to set the limit in the range $10^{-13}< \sin^2 2\theta_S<10^{-5}$ for
$m_S = (0.6 - 40)$ keV.

\subsection{CMB-bound}

If the sterile neutrinos decay producing light neutrinos between the
active neutrino decoupling time, $\tau_\nu \sim 1$ s, and the photon
decoupling time, $\tau_{\rm rec}$, this would increase the energy
density of relativistic particles at $t \lsim \tau_{\rm rec}$.  The
density is described by the effective number of neutrinos, $N_\nu$.
That, in turn, affects the CMB angular power spectrum (acoustic
peaks). The bounds on $N_\nu$ from observations were substantially
improved during the last 5 years: $N_\nu<13$ at 95\% CL
(BOOMERanG/MAXIMA)~\cite{boomerang}, $N_\nu<8.3$ at 90\% CL (WMAP
data)~\cite{barger}, $N_\nu<6.8-7.1$ at 95\% CL (WMAP and
LSS)~\cite{crotty,pierpaoli,hannestad1}, $N_\nu<5.4$ at 95\% CL (WMAP,
LSS and type Ia supernova data)~\cite{hannestad2}.  

With the help of considerations in Ref.~\cite{raffelt} and
~\cite{fuller}, we get the ``CMB'' limit on the sterile neutrino
parameters as a function of $N_\nu$, which can be parametrized as

\begin{equation}
  \left( \frac{m_S}{\rm keV} \right)^4=3\times 10^{33}\,(N_\nu-3)^{-2.87}\; \sin^2\theta_S\, .
\label{eq:cmb}
\end{equation}
Combined analysis of the cosmological data on LSS, supernovas and the
CMB including 3 years result from WMAP \cite{WMAP-3} allows to put the
bound $N_\nu < 3.74$ at 95\% CL.  
Using this results and Eq.~(\ref{eq:cmb}) we find the bound shown 
in Figs.~\ref{fig1}-\ref{fig3}. This limit is not valid above the line
$\tau_S < \tau_\nu \sim 1$~s, which for $\sin^2\theta_S = 10^{-12}$
corresponds to $m_S = (400 - 500)$ MeV.

In the future, the PLANCK mission~\cite{planck1,planck2} will allow to
strengthen the bound down to $N_\nu<3.2$, while according to
Ref.~\cite{planck2}, the CMBPOL mission can achieve $N_\nu<3.05$. This
will further expand the excluded region, in particular to larger
values of $m_S$.

\subsection{BBN bound} 

Apparently for one sterile neutrino the limit on the effective number
of additional degrees of freedom during the epoch of Big Bang
Nucleosynthesis (BBN), $\Delta N_\nu=1$, does not provide any bound since
at most the equilibrium concentration of $S$ can be produced 
in the scenarios under consideration.  The
limit $\Delta N_\nu \leq 1$ becomes relevant in the case of more than
two sterile states.

For the low values of $m_S$ we use the limits from ~\cite{villante}
which for $\Delta N_\nu=1$ can be parametrized as 
\be 
m_S\, \sin^2\theta_S =  1.4 \; {\rm meV}, \quad
{\rm for} \quad \nu_e, 
\ee
\be 
m_S \, \sin^2\theta_S = 1.0\; {\rm meV}, \quad {\rm for} \quad  
\nu_\mu, \nu_\tau. 
\ee

For the high masses, $m_S = 10 - 200$ MeV,  the exclusion region 
has the shape of a parallelogram~\cite{dhrs}.
The right boundary is basically given by
$$
\sin^2 \theta_S \left( \frac{m_S}{\rm keV}\right)^3 = 1.25 \cdot 10^{11}, 
$$
which is the condition for the heavy neutrinos to be relativistic at
decoupling so their number density is not Boltzmann suppressed and they
can have an impact on BBN.
The left boundary,  
$$
\left(\frac{m_S}{\rm keV}\right)^2 \sin^2 \theta_S < 6 \times
10^{-2}, 
$$ 
can be simply understood 
as  the condition that the energy density of the
sterile neutrino is smaller (including also an entropy dilution factor 5) 
than the energy density of one light neutrino species at BBN.  

\subsection{Supernova neutrino bound}

Two different bounds follow from observation of the antineutrino signal 
from SN1987A~\cite{1987a}. One is from $\bar\nu_e$-disappearance 
and the other from  star cooling. 

1) The resonance conversion $\bar \nu_e \to \bar S$ can occur in the
central regions of the star due to change of sign of the matter potential.
That happens for the active-sterile system when the relative electron
number density is $Y_e \sim 1/3$.  For the mass range $m_S \sim (1 - 100)$
eV the adiabaticity condition can be fulfilled if $\sin^2\theta_S >
10^{-5}$.  The adiabatic conversion leads to strong suppression of the
$\bar\nu_e$- flux \cite{orlando}. So, the observation of the
$\bar\nu_e$-signal from SN1987A gives the bound on the oscillation
parameters.  In Figs.~\ref{fig1}-\ref{fig2} we show the updated
results obtained in Ref.~\cite{vissani}.

2) For large masses, $m_S$, the $(\nu_a \to S)$ oscillations as well
as scattering lead to production of sterile neutrinos in the core of
the collapsing star. If these neutrinos escape the core, they can lead
to substantial energy loss.  This, in turn, will shorten the neutrino
burst and the energy released in $\bar\nu_e$ will be smaller
~\cite{hansen,maalampi}.

Normal duration of the SN1987A neutrino burst excludes significant
cooling effect which puts the bound on $S$ parameters~\cite{dhrs}
shown in Figs.~\ref{fig1}-\ref{fig3}.  Notice that the upper bound,
$\sin^2 \theta_S \lsim 10^{-10}$, is slightly different for $\nu_e$
and $\nu_\mu$ and $\nu_\tau$.  If mixing is sufficiently large, the
sterile neutrinos will be trapped inside the core, and the energy-loss
argument is not applicable.  This gives the lower bound of the
excluded region $\sin^2\theta_S \lsim 10^{-2}$.  The mass range is
restricted by the condition that $S$ is produced inside the core.

Future detection of high statistics neutrino signal from a 
Galactic supernova will allow to put stronger bounds.  
Indeed, another resonance in $\nu_e - S$ can occur in the outer regions
of the star with normal chemical composition. 
The adiabaticity condition can be written as 
\be
\gamma =  10^2 \sin^2 \theta_S  
 \left(\frac{m_S}{\rm 10^{-3} {\rm keV}}\right)^{3.3} \gg 1 
\ee
which in fact is stronger than the cosmological bound in the range
(0.01 - 0.1) keV.  The adiabatic $\nu_e - S$ conversion leads to the
disappearance of the $\nu_e$-neutronization peak, and the modification
of the signal during the cooling stage. In particular, in the case of
normal mass hierarchy the electron neutrino flux at the Earth will be
$F({\nu_e}) = F^0({\nu_\mu})$ in the case of large 1-3 mixing and
$F({\nu_e}) = \cos^2\theta_{12} F^0({\nu_\mu})$ for very small 1-3
mixing~\cite{orlando}.

In the eV mass range, sterile
neutrinos can drive non negligible $\nu_\mu \to \nu_e/\bar \nu_\mu
\to \bar\nu_e,\bar\nu_\tau$ conversions at TeV energies by the MSW
effect that can be constrained by future IceCube data~\cite{hpz}.

\subsection{Laboratory bounds}

The laboratory  bounds  are typically much weaker than the 
astrophysical and cosmological ones. They are, however, more 
robust and  turn out to be the main bounds if, for some reason 
(see sec.~\ref{sec:avoid}), the cosmological and astrophysical 
limits become inapplicable.

1). {\it The $0\nu\beta\beta$-decay.} 
Introduction of active-sterile mixing (described by 
$m_{\alpha S}$) does not change the $m_{ee}$ element of the 
whole $4 \times 4$ mass matrix (\ref{total-m}). Therefore
for light $S$: $m_S \ll 1/r_N$, where $r_N$ is 
the typical size of  nuclei, the rate of  $0\nu\beta\beta$-decay 
in the lowest approximation is determined by  $m_{ee}$  
and the effect of $S$ is strongly suppressed.    

However, the effect of $S$ on the  $0\nu\beta\beta$-decay  
increases with $m_S$. Let us consider this in more details.  
The $0 \nu \beta  
\beta$ decay amplitude, $A^{(S)}$, has two contributions associated to $S$: i)
from the induced mass, $(m_I)_{ee}$, and  exchange of light neutrinos,   
and ii) from exchange of $S$ and its mixing with $\nu_e$. So, the amplitude
of the decay can be written as
\be 
A^{(S)} \propto \frac{(m_I)_{ee}}{ \bar{q}^2 - m_{\nu}^2} + 
\frac{m_S \sin^2 \theta_{eS}}{\bar{q}^2 - m_S^2}, 
\label{abeta}
\ee where $\bar{q} \sim 1/r_N$ is the effective momentum of the
exchanged neutrino. 
As it can be inferred from the calculation presented in~\cite{nuless},
$\bar{q} \approx 100$ MeV.  In the denominator of (\ref{abeta})
the parameter $m_{\nu}$ is the effective mass of light neutrinos and since 
$m_{\nu} \ll m_S$ it can be neglected.  Taking into account that the 
induced mass $(m_I)_{ee} = - m_S \sin^2 \theta_{eS}$ we can write, 
according to Eq.~(\ref{abeta})), the total contribution of $S$ to the 
effective  Majorana mass as 
\be
m_{ee}^{(S)} = m_S \; \sin^2 \theta_{eS} \, \displaystyle  \left \vert 1 -
  \frac{\bar{q}^2}{\bar{q}^2 - m_S^2}\right \vert \, .
\label{effmass}
\ee
If $m_S$ is negligible the two contributions cancel each other leading 
to  zero effect of $S$. If $m_S^2 \ll \bar{q}^2$, we find from (\ref{effmass}) 
\be
m_{ee}^{(S)} =  \sin^2 \theta_{eS} \frac{m_S^3}{\bar{q}^2}\, .
\label{lowmass}
\ee
So, for light $S$, $m_S \ll 1/r_N$, $m_{ee}^{(S)} \rightarrow 0$ and  the 
total  effective Majorana mass  is determined  by
the $ee$-elements of the active neutrino mass matrix 
${\bf (m_a)}_{ee}$, in agreement with consideration in terms of $4 \times 4$ matrix.  

\begin{figure}[htb]
\epsfxsize=14cm
\begin{center} 
\leavevmode 
\epsffile{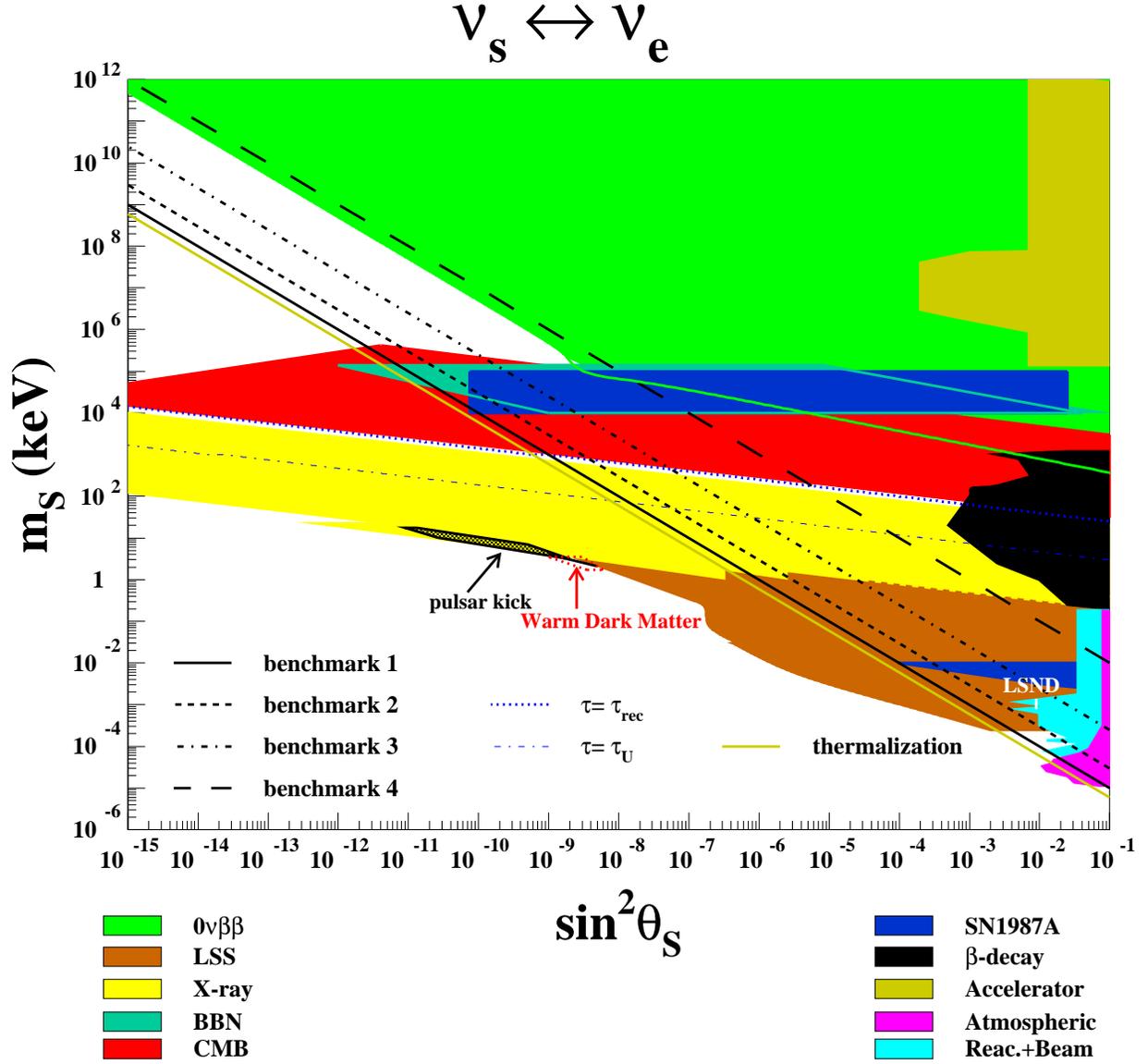}
\end{center}
\vglue -3.5cm
\caption{ The benchmark lines Eqs.~(\ref{small})-(\ref{leadA}) and
  (\ref{bench4}) versus the current astrophysical, cosmological and
  laboratory bounds on $\nu_S -\nu_e$ mixing as described in the text.
  The colored regions are excluded in each case.  The
  ``thermalization'' line and the two decay lines $\tau_S = \tau_{\rm
    rec}$ and $\tau_S = \tau_U$ are also shown.
 We show also the allowed regions for the warm dark matter and 
the LSND (3+1) as well as the region that could explain pulsar velocities.
}
\label{fig1}
\end{figure}

For $m_S^2 \gg \bar{q}^2$, the second term in (\ref{effmass}) 
can be neglected  and the Majorana 
mass is given by the induced contribution (as in the usual see-saw mechanism): 
\be
m_{ee}^{(S)} =  \sin^2 \theta_{eS}{m_S} \, .
\label{highmass}
\ee In Fig.~\ref{fig1} we show the excluded region of the $S$
parameters which corresponds to the upper bound $m_{ee} < 0.5$ eV
obtained from studies of $0\nu\beta\beta$-decay of $^{76}$Ge
\cite{heidelberg}. We assume that there is no cancellation between $S$
contribution and ${\bf (m_a)}_{ee}$.  In general, cancellation reduces
the excluded region. However, in our context, the cancellation can not
be strong.  Indeed, according to our assumption (Sec. 2.1) the
original active neutrino mass terms are below 1 eV. Therefore maximal
contribution to the effective Majorana mass of electron neutrino is
about 1 eV, and consequently, only the order of 1 eV contribution from
new neutrino state can be cancelled. That corresponds to the long-dashed
line in Fig 1.

As follows from  Fig.~\ref{fig1} the double beta  
decay limit becomes relevant for high masses $m_S > 100$ MeV.

Future neutrinoless double beta decay experiments~\cite{fut0nu} 
with sensitivity  down to $m_{ee} < (0.01 - 0.03)$ eV
will improve the bound on $\sin^2 \theta_S$  by a factor 10-30. \\

2) {\it The $\beta$-decay}. The region 
$m_S \sim (0.1 - 10^{3})$ keV and $\sin^2\theta _S \lsim 10^{-3}-10^{-2}$ for 
$\nu_S- \nu_e$ mixing  (Fig.~\ref{fig1}) is  excluded
by the negative results of searches of  kinks  
in the energy spectra of nuclear $\beta$-decays~\cite{beta}.\\

\begin{figure}[htb]
\epsfxsize=16cm
\begin{center} 
\leavevmode 
\epsffile{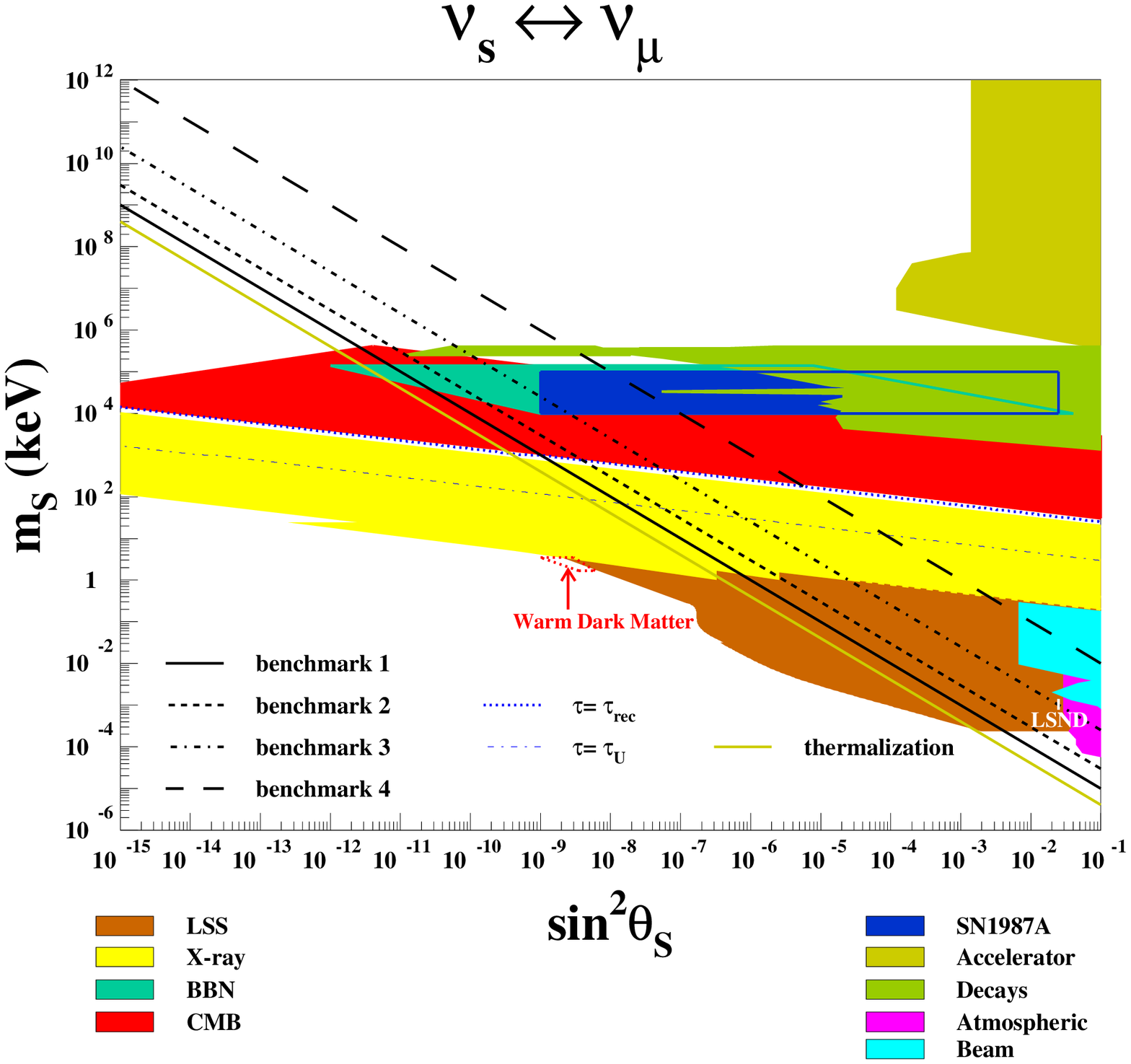}
\end{center}
\vglue -1cm
\caption{Same as Fig.\ref{fig1} but for $\nu_S -\nu_\mu$ mixing.}
\label{fig2}
\end{figure}

\begin{figure}[htb]
\epsfxsize=16cm
\begin{center} 
\leavevmode 
\epsffile{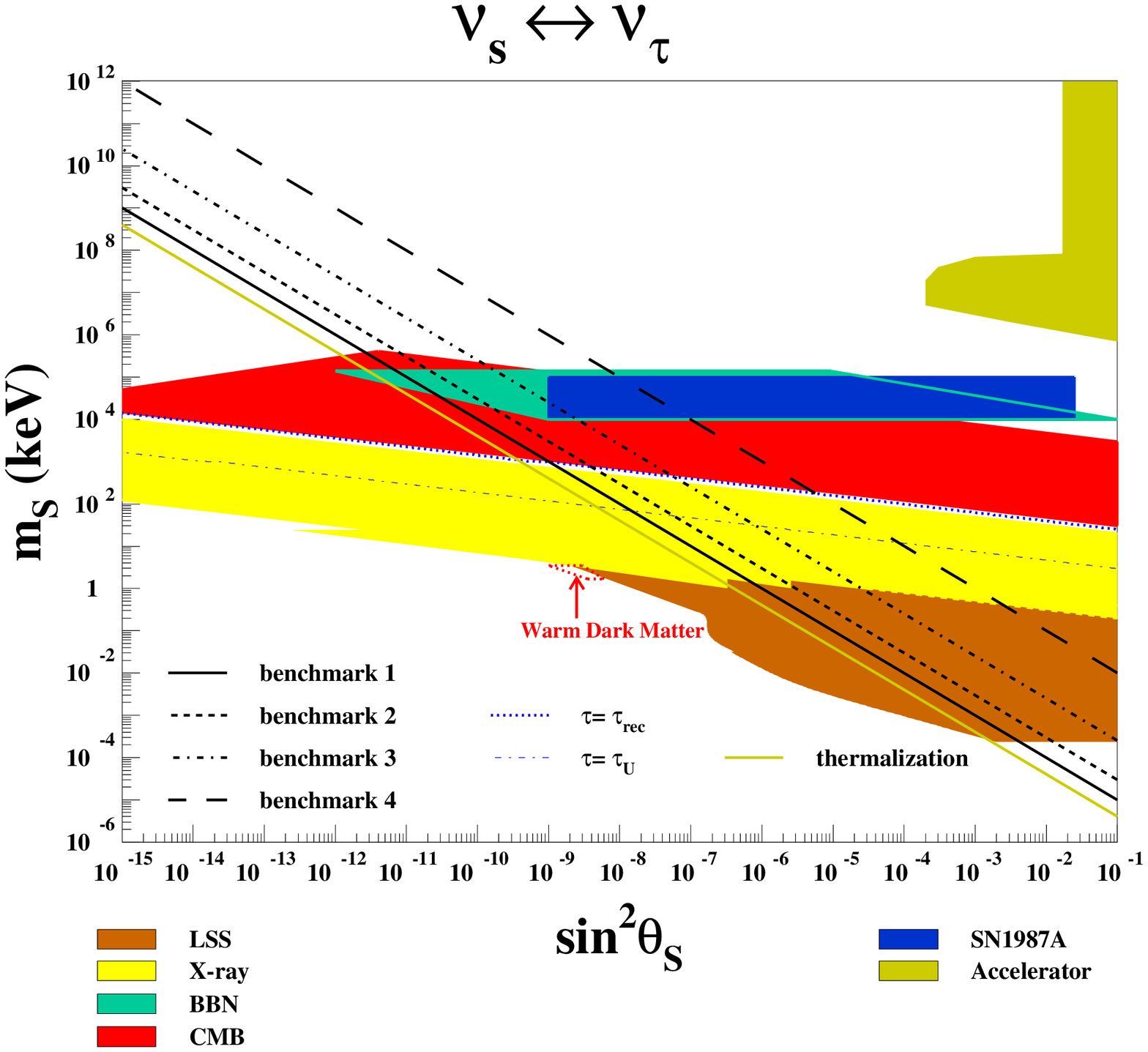}
\end{center}
\vglue -1cm
\caption{Same as Fig.\ref{fig1} but for $\nu_S -\nu_\tau$ mixing.}
\label{fig3}
\end{figure}

3) {\it Meson decays} can have contributions from sterile neutrinos
which can modify the energy spectra of their decay products.
In the $\nu_S - \nu_e$ channel, the best limit comes from
precision measurements of the energy spectrum of $e^+$ in the decay
$\pi^+ \to e^+ \nu_e$~\cite{britton}: $\sin^2
\theta_S < 10^{-7}$  for  $ m_S = (50 - 130)$ MeV. 
In the $\nu_S -\nu_\mu$ channel, 
the bounds come from studies of the spectra of $\pi^\pm$ and $K^\pm$ 
decays at  accelerators and in
the atmosphere. The $90\%$ CL excluded region 
(labeled as ``Decays'' in Fig.~\ref{fig2})  is taken from 
Ref.~\cite{kusenko}. It  
is the only bound which can compete with the cosmological and
astrophysical limits in the range of large masses $m_S \sim (0.03 -
0.3)$ GeV.

In the near future, the studies of pion, muon and kaon decays at
MiniBooNE~\cite{miniboone}, MINOS~\cite{minos} and K2K~\cite{k2k}
experiments are expected to reach sensitivities of a few$\times
10^{-7}$ or less for the mixing if $m_S \sim 100$ MeV~\cite{kusenko}.

Oscillations of active to sterile neutrinos lead to the suppression of
the neutral current interaction rate. In this connection, it has been 
proposed to study interactions of neutrinos from pion decays at rest
with the superallowed neutral current reaction $\nu_x~ ^{12}$C
$\rightarrow \nu_x~ ^{12}{\rm C}^*$~\cite{garvey}. The sensitivity of
such an experiment can reach $\sin^2\theta_S \sim 10^{-2}$ for $m_S
\sim 1 $ eV. \\

4) {\it The atmospheric neutrinos and K2K.} 
The combined analysis of the  SuperKamiokande~\cite{sk} and 
MACRO~\cite{macro}  results as well as the data from accelerator 
experiment K2K  leads to the following $90\%$ CL. 
bounds~\cite{vissani}:    for the $\nu_{e} - \nu_S$ 
channel $\sin^2\theta_S< 10^{-2}$, in the
interval $m_S=(0.01-0.1)$ eV (see Fig.~\ref{fig1});   for the
$\nu_{\mu} - \nu_S$ channel  $\sin^2\theta_S< 0.04$ 
in the range $m_S = (0.1 - 1)$ eV (see
Fig.~\ref{fig2});   for the $\nu_\tau - \nu_S$ mixing  
$\sin^2\theta_S< 0.1$. 

If the cosmological bound is invalid,  
the atmospheric bound becomes also relevant for larger masses,   
$m_S = (1 - 10)$ eV. 

Sterile neutrino oscillations for $m_S \sim$ few eV have 
the oscillation length comparable to the Earth's radius at 
$E_\nu \sim $ 1 TeV. 
In the future, IceCube~\cite{icecube} detection of atmospheric neutrinos 
can probably   
bring the limit on $\nu_e - \nu_S$ mixing down to 
$ \sim 10^{-2}$ for $m_S \sim $ few eV~\cite{hpz}. 

It has been suggested in Ref.~\cite{gelmini}  that showers generated by 
ultra-high energy sterile neutrinos ($E_\nu \sim 10^{6}-10^{12}$ GeV)
with $\sin^2 \theta_S \sim 0.01-0.1$, may be distinguished from the ones 
generated by active neutrinos in experiments such as EUSO~\cite{euso} or OWL~\cite{owl}, 
using the Earth as a filter, as proposed in Ref.~\cite{barbot} for 
neutralino showers.  
The mass range of $m_S$ that can be  
tested by these experiments is determined by the mechanism 
of their production. \\

5) {\it Reactor neutrino experiments} can set limits on the 
mixing $\bar \nu_e - \bar \nu_S$ by comparing expected with observed 
$\bar \nu_e$ flux. 
Bugey~\cite{bugey}, CHOOZ \cite{chooz} 
and Palo Verde~\cite{palo} provide the main bound
on $\bar \nu_e - \bar \nu_S$ mixing, $\sin^2 \theta_S \lsim
0.01$  ($90 \%$  CL.), in the range $m_S = (0.3 - 1)$ eV. 

The forthcoming  reactor experiment Double-CHOOZ and the proposed 
projects Kaska, Braidwood and Angra~\cite{whitepaper}, may be able to 
reach $\sin^2 \theta_S \sim 5 \cdot 10^{-3}$  for 
$m_S = (0.03 - 0.7)$ eV~\cite{theta13}. \\

6) {\it Accelerator bounds.}  The accelerator oscillation experiments
CDHS~\cite{cdhs}, CCFR~\cite{ccfr}, NOMAD~\cite{nomad} and
CHORUS~\cite{chorus} as well as LSND~\cite{lsnd},
KARMEN~\cite{karmen}, provide with the bound on $\nu_{\mu} - S$ mixing
in the range low mass range, $m_S > (1 - 100)$ eV. In particular,
$\sin^2 \theta_S < 7 \cdot 10^{-3}$ (90\% CL ) for $m_S > 10$ eV (see
region labeled ``Beam'' in Fig.\ref{fig2}).

In Figs.~\ref{fig1} we show the forbidden region of the
parameters (labeled ``Reac.+Beam'') from the combined analysis of
these accelerator beam and reactor experiments Ref.~\cite{vissani}.
This result excludes the induced mass as the origin of the dominant
structure in the low $m_S$ domain.

In the high mass range, $m_S > 0.1$ GeV mixing with a heavy $S$ leads
to an effective violation of lepton universality and appearance of the
flavor changing neutral currents.  The non-observation of these
effects in experimental data permitted the authors of Ref.~\cite{nardi} 
to set the 90\% CL limits, shown in Figs.~\ref{fig1}-\ref{fig3}: 
$\sin^2\theta_S<7.1 \cdot 10^{-3}$ ($\nu_S- \nu_e$),  
valid in the range $m_S > 0.14$ GeV $(m_\pi$), 
$\sin^2\theta_S<1.4 \cdot 10^{-3}$ ($\nu_S - \nu_\mu$),  
valid in the range $m_S > 1.115$ GeV ($m_\Lambda$), and
$\sin^2\theta_S<1.7 \cdot 10^{-2}$ ($\nu_S - \nu_\tau$), 
valid in the range $m_S > 1.777$ GeV ($m_\tau$).
Also searches for a singlet neutral heavy lepton at LEP~\cite{l3}, 
have allowed to exclude $\sin^2\theta_S$ down to $10^{-4}-10^{-1}$ 
depending on the channel, in the range $0.4<m_S/{\rm GeV}<90$ at 95\% CL.

Accelerator experiments sensitive to oscillations with $\Delta m^2 >
\Delta m^2_{32}$ will be able to set stringent bounds in the low mass
range.  The MINOS detector~\cite{minos} can use the neutral current to
charged current ratio to probe $\nu_\mu -\nu_S$ and $\nu_\tau - \nu_S$
mixing down to $\sin^2\theta_S \sim 10^{-2}$~\cite{vissani}.
Sensitivity to $\nu_{\mu}- \nu_S$ mixing at the level $\sin^2 \theta_S
\sim 10^{-3}$ for $m_S = (0.03 - 0.3)$ eV can be reached by
T2K~\cite{t2k}, NO$\nu$A~\cite{nova} and future $\nu$-factories
looking for $\nu_\mu, \bar \nu_\mu$ disappearance.

Forthcoming  $pp$  and planned $e^+e^-$  collider experiments can 
expand the the excluded region  of  $S$ parameters to larger  masses. 
According to Ref.~\cite{ali}, the search for same-sign
dilepton production mediated by a sterile neutrino $S$ in 
$pp \rightarrow l^+ l^{'+} X$ with $l,l'=e,\mu,\tau$, may be used to
constrain $\sin^2 \theta_S \lsim 10^{-2}$ for $m_S= (0.1-2)$ TeV 
at the Large Hadron Collider at CERN, provided a large integrated 
luminosity becomes available.
It has been shown~\cite{ilc} that at an International Linear
Collider with a center of mass energy $\sqrt{s}=$ 500 GeV, one can
look for single heavy $S$ production through $e^+e^-\rightarrow S \nu
\rightarrow l W \nu$ with $l=e,\mu,\tau$.  This has a sensitivity down to
$\sin^2 \theta_S < 7 \cdot 10^{-3}$ for $m_S = (200 - 400)$
GeV. The same reaction at a future Compact Linear Collider 
operating with $\sqrt{s}=$ 3 TeV could be used to limit
$\sin^2 \theta_S < (2-6) \cdot 10^{-3}$ for $m_S=1-2$ TeV~\cite{clic}.

Notice however, that these high mass bounds are essentially irrelevant
in our context.  Indeed, for $m_S > 1$ GeV and $\sin^2 \theta_S >
10^{-3}$, the induced mass is $m_I > 1$ MeV. This means that the
elements of the original active neutrino mass matrix should also be
large $m_a > 1$ MeV, and it should be extremely strong cancellation of
the original and induced contributions to obtain phenomenologically
acceptable masses of light neutrinos: $(({\bf m_a})_{ij} - ({\bf
  m_I})_{ij})/ (({\bf m_a})_{ij} < 10^{-7}$.
 
Inversely, discovery of the $S$ with relatively large mixing 
in high energy collisions will testify against the approach 
developed in this paper.

\section{Induced mass versus  direct mixing effects}

\subsection{Bounds on induced mass}

Confronting the lines of constant 
induced masses (\ref{small}), (\ref{sublead}) and (\ref{leadA})
with cosmological astrophysical and laboratory  bounds 
in the $\sin^2\theta_{S}- m_S$ plane we can 
conclude on the relative importance of the direct and indirect  effects.

There are two regions of parameters in the $\sin^2\theta_{S}- m_S$
plane, where the induced masses are more important than the effects of
direct mixing.  That is, in these regions substantial induced masses
for the active neutrino mass matrix are not excluded by the existing
bounds.

1) {\bf High mass region:} $m_S \gsim 300$ MeV and $\sin^2\theta_{S} \lsim
10^{-9}$.  This region is restricted essentially by the CMB
bound, meson decays and SN1987A cooling.  Future measurements can
probably improve the bounds by about 1 order of magnitude from below.

Only for the $\nu_e - \nu_S$ channel, the neutrinoless double beta decay
can probe the whole high mass region. Present bounds correspond
essentially to the dominant contribution in the case of the degenerate
spectrum.

For the other mixing channels, it is the induced mass which gives the
bound on the parameters of the sterile neutrinos. Indeed, assuming
that there is no strong cancellation of elements of the original
active neutrino mass matrix ${\bf m_a}$ and ${\bf m_I}$, and taking
the largest elements of matrices in Table 1, we can write the bound
\be 
\sin^2\theta_{S} m_S \lsim m^{\rm exp} \sim (0.5 - 1) \;{\rm eV},
\label{bench4}
\ee 
which is clearly comparable to the neutrinoless double beta
decay bound but now valid for all channels. This can be viewed as 
the forth benchmark line shown in Fig.\ref{fig1}-\ref{fig3}.

Sterile neutrinos in this range can play some role in  leptogenesis
and the generation of the baryon asymmetry 
in the Universe~\cite{ars,Asaka:2005pn}.

Clearly contribution from this region is out of our control and this 
creates ambiguity in the implications  of the  mass and mixing results. 

It is beyond the scope of this paper to discuss the possible origins
of $S$ with such a small mixing.  Still one possibility looks rather
interesting: if $m_S = (10^2 - 10^6)$ GeV (region where one may expect
singlets related, {\it e.g.}, to SUSY breaking), the required mixing is
$\sin^2\theta_{S} = 10^{-15} - 10^{-12}$. The latter can be related to
the existence of a new ``intermediate'' scale $M = m_S/ \sin
\theta_{S} \sim (10^{10}-10^{12})$ GeV.

If the interpretation of the  LSND result in terms of oscillations 
in (3 + 1) scheme is confirmed, that would imply existence of  
the sterile neutrino(s) with mass  $(0.5  - 5)$ eV with mixing parameters  
$\sin^2 \theta_S \sim 0.02$ (see ~\cite{goswami} for recent analysis). 
The  corresponding induced mass equals  
$m_I = (10 - 100)$ meV and therefore the effects of LSND neutrino  
on the active neutrino mass matrix is strong and can not be considered as  
small perturbation.

2) {\bf Low mass window:} $m_S \sim (0.1 - 0.3)$ eV and
$\sin^2\theta_{S} = 10^{-3} - 10^{-1}$.  This window is essentially
closed for all the channels if one takes the BBN bound $\Delta N_{\nu}
< 1$. In this case, one has the bound on the induced mass $m_I < 1$
meV. That can produce some effect on observables in the case of
degenerate or inverted hierarchy spectrum.  From above this region is
restricted by the LSS bound on neutrino mass.

If $\Delta N_{\nu} = 1$ is allowed, there is no BBN bound.  Bounds from
other effects strongly depend on flavor.  For the $\nu_e - \nu_S$
channel, the reactor and atmospheric neutrino bounds essentially
exclude the dominant contribution from $m_I$ but still allow for
sub-dominant effects.  For the $\nu_\mu - \nu_S$ channel, the bound is
given essentially by the atmospheric neutrinos and larger region of
mixings is allowed. In particular, for $m_S \sim 0.25$ eV the mixing
$\sin^2\theta_{S} = 0.04$ is not excluded leading to $m_I \sim 10$
meV. The latter is close to the dominant contribution.  For the
$\nu_\tau - \nu_S$ channel, the atmospheric bound is weaker and
dominant contributions from $m_I$ are allowed:
$m_I \sim (30 -  250)$ meV.\\
  
In the rest of the $m_S$ region, $m_S = (10^{-3} - 10^{5})$ keV, effects
of the direct mixing dominate over the induced matrix effects.
In this range the induced masses  
\be 
m_I \lsim 4 \cdot 10^{-2} \; {\rm meV},
\ee
can produce only very small corrections to the active neutrino mass matrix. 
In  the interval $m_S = (1 - 10^{4})$ keV the bound is even stronger: 
\be
m_I \lsim 10^{-2} \; {\rm meV}. 
\ee

If however the cosmological/astrophysical bounds are absent for some
reason, a large range of parameters becomes allowed and the induced
masses can reproduce the dominant structures of the active neutrino
mass matrix in the whole range of $m_S$.

In Fig. 1 - 3 we show also regions of parameters which 
correspond to certain positive indications of the existence of sterile 
neutrinos: (1) overlapping regions of the warm dark matter and pulsar 
kick, (2) the LSND spot. 

In the scenario \cite{Asaka:2005pn} two neutrinos have  masses 
$m_S \gsim 1$ GeV and their mixing is responsible for the 
mass matrix of light active neutrinos: ${\bf m_a} = {\bf m}_I$. 
The third sterile neutrino with mass  $m_S \sim 1$ keV and 
mixing $\sin^2 \theta_S \sim 10^{-9} -  10^{-8}$ can contribute 
substantially to warm dark matter (WDM) and  explain pulsar kicks.  

Apparently the WDM  and pulsar kick regions are far below  the 
benchmark lines and therefore the corresponding induced masses are 
negligible. Furthermore, the regions are disfavored by the 
cosmological and astrophysical observations. 
The WDM scenario can be recovered 
if the mixing is smaller then that indicated in the plot  
and some additional mechanism of generation  
of sterile neutrinos exists apart from mixing with active neutrinos. 

In contrast, the LSND spot is in the range where the induced masses are of 
the order of dominant mass structures. So mixing with sterile neutrino 
can  not be considered as small perturbation of the original active 
neutrino structure. The LSND spot is essentially excluded by 
the cosmological data unless some new physics is added.

\subsection{Avoiding bounds}
\label{sec:avoid}
Let us consider various possibilities which allow one to circumvent the 
bounds obtained in the previous section and therefore to 
open a possibility for strong effects of the induced matrix even 
for low  $m_S$.

1) It has been shown~\cite{gelmini} that in a cosmological scenario
with low reheating temperature at the end of inflation, $T_R << 100$
MeV, experimental bounds on the active-sterile neutrino mixing are
relaxed.  In particular, the sterile neutrino required to explain the
LSND result is allowed in this scenario.  Also the influence of a
non-negligible primordial lepton number asymmetry has impact on the
sterile production rate since in this case neutrinos are produced
resonantly with a non-thermal spectrum~\cite{shi2,fuller,bell}.
According to Ref.~\cite{fuller}, for $L=10^{-3}$ the LSS limit
presented in Figs.\ref{fig1}-\ref{fig3} will change very little,
while for $L= 0.1$ the limit on $\sin^2\theta_S$ can increase 
by 1 or 2 orders of magnitude depending on $m_S$.

2) Another possibility can be related to the origin of the mass of
$S$ itself.  Recall that $S$ may not be related to right handed
neutrinos and usual family structure.  Let us assume that $S$ has a 
``soft mass'' generated by the medium dependent VEV of some new scalar field
$A$: $m_S = \lambda \langle A \rangle$.  The VEV can be proportional
to the number density of the active neutrinos, $n_{\nu}$: $\langle A
\rangle \propto n_{\nu}$~\cite{fardon}.  In this case 
\be 
m_S = m_S^0 (1+z)^3, 
\ee 
where $m_S^0$ is the mass in the present epoch.  If the
mixing mass is generated by the usual Higgs VEV, which does not change
with time, we  find that the induced contribution to the light mass
and the active-sterile mixing  decrease back in time: 
\be
m_{I} = \frac{m_{iS}^2}{ m_S^0 (1+ z)^3} = m_{I}^0 (1 + z)^{-3}, 
\ee
\be 
\sin \theta_S = \frac{m_{iS}}{ m_S^0 (1+ z)^3} = \sin \theta_S^0
(1 + z)^{-3}.  
\ee 
Here $m_{I}^0$ and $\theta_S^0$ are parameters at 
the present epoch.  The combination we plot in Figs.~\ref{fig1}-\ref{fig3} 
changes as 
\be
\sin^2 \theta_S m_S = \frac{\sin^2 \theta_S^0 m_S^0}{(1+ z)^3}.  
\ee
The lines of constant induced mass shift with $z$ to the left - to
smaller masses and mixings.  This means that in the past all
cosmological bounds where satisfied.

Already at the recombination epoch the mass of  sterile neutrino
becomes of the order $10^4$ GeV for the present mass 
$m_S^0 \sim 1$ eV and the mechanism of oscillation
production does not work. Essentially, in this scenario the
sterile neutrinos are not produced in the Early Universe and  their 
concentration is negligible.
The astrophysical and cosmological bounds we have discussed are not
applicable.

3) If $S$ interacts with a massless or low-mass
Majoron~\cite{majoron} $\phi$, it can decay invisibly as $S \to \nu
\phi$ or annihilate (see similar mechanism for the active 
neutrinos in \cite{bbd}). If this decay is fast enough, 
$\tau_S << 1$ s, in principle, 
all astrophysical and cosmological bounds could be evaded.
See \cite{palomares} for the recent similar analysis.  
In the mass range $m_S \sim 1$ keV such a fast decay can be achieved 
for the scalar coupling $g \sim 10^{-8}$.

In the low mass range some restrictions  
on  the off-diagonal couplings 
$g \nu_a S \phi$ can be obtained from their effect on the free-streeming
condition for active neutrinos. The latter should manifest itself
in the precision measurements of the CMB acoustic peaks.
One can use these bounds also for sterile neutrinos. 
In \cite{raffelt2} a limit on the active neutrino  coupling $g\lsim 1 \cdot 
10^{-11} (50 \rm \; meV/m_S)^2$ was obtained.  
According  to Ref.~\cite{bell2},  the limit is less severe 
and  is absent if couplings with 
different active neutrinos are different,  
{\it e.g.},  S has large coupling with $\nu_{\tau}$ only.

\section{Conclusions}

The main conclusions of this paper can be summarized as follows.

1) Mixing of the active neutrinos with sterile neutrinos, singlets of
the SM symmetry group, generates an induced mass matrix of active
neutrinos which can be the origin of peculiar properties of the lepton
mixing and neutrino mass spectrum. It opens an alternative possibility 
to understand possible new symmetries in the neutrino sector.  In this
way, one can explain the substantial difference of mixing patterns of
quarks and leptons.

Depending on masses and mixings of $S$, the induced active neutrino
masses can be the origin of the dominant or sub-dominant structures of
the neutrino mass matrix. For instance, the tri-bimaximal mixing can
originate from the induced contribution.

2) Apart from modification of the mass matrix of active neutrinos,
there are direct mixing effects of $S$ which can be observed in
cosmology, astrophysics and laboratory experiments.

The importance of the direct and induced effects depends on the range
of $S$ parameters considered.  For $m_S \gsim 300$ MeV the induced
effect dominates. The induced masses can reproduce the dominant
structures of the active mass matrix. The direct mixing effects are
negligible.  In the interval $m_S \sim (10^{-3} - 10^{5})$ keV the
direct mixing effects dominate: the astrophysical and cosmological
consequences of mixing are more important putting strong upper bounds
on the induced mass. So, the latter can be neglected in the mass
matrix of active neutrinos.  In the narrow window $ m_S\sim
  (0.1 - 0.3)$ eV the two effects are comparable. For the BBN bound
$\Delta N_{\nu} < 1$ the induced matrix can produce only small
effects. If one additional neutrino is allowed by BBN, the induced
masses can generate the sub-leading structures or even be comparable
to the values of dominant mass matrix elements.

3) New physics effects can relax or even lift the cosmological and
astrophysical bounds thus opening a possibility to generate large
induced matrix in the whole range of masses $m_S > 0.01$ eV.
Here interesting possibilities to notice are: cosmological scenarios
with low reheating temperature, fast decay of $S$ into a Majoron and
neutrino, and  the possibility of a soft mass $m_S$ which varies 
with time, {\it a la}  Mass Varying Neutrino scenario.

\section*{Acknowledgments} 

We are grateful to A. Slosar for proving us with bounds on the energy
density of sterile neutrinos in the whole range of masses.  We thank
U.~Seljak for useful discussions and G.~G.~Raffelt, S.~Pastor,
A.~Kusenko, M.~Shaposhnikov and N.~Bell for valuable correspondence.
This work was partially supported by Funda\c{c}\~ao de Amparo \`a
Pesquisa do Estado de S\~ao Paulo (FAPESP) and Conselho Nacional de
Ci\^encia e Tecnologia (CNPq).  R.Z.F. acknowledges the Phenomenology
Institute of the University of Wisconsin at Madison where the final
part of this work was done.

\vglue -6cm
\vglue -1cm
\vglue -1cm
\begin{landscape}
{\tiny
\begin{table}
\begin{center}
\begin{tabular}[t]{|c|c|c|c|}
\hline
{\rule[-3mm]{0mm}{12mm}\bf Case} & {\bf Best Fit} & {\bf Experimentally allowed at 1 $\mathbf\sigma$} & {\bf Free CP phases} \\
{\rule[-4mm]{0mm}{8mm}} & $ \delta=\lambda_3=0$ & $\delta=\lambda_3=0$ & \\ 
\hline
{\rule[-9mm]{0mm}{20mm}\bf Normal} & $\small\left(\begin{array}{rrr} 
3.2 \hfill& 6.0 & 0.6\\
 & 24.8 & 21.4\\
 &  & 30.7
\end{array}\right)$ & $\small \left(\begin{array}{rrr} 
2.5-5.0 & 2.7-9.8 & 0.-5.1\\
 & 19.9-30.3 & 18.1-22.9\\
 &  & 24.5-34.0
\end{array}\right)$ & $\small \left(\begin{array}{rrr} 
0.3-5.0 & 0.-10.8 & 0.-11.1\\
 & 12.7-30.9 & 18.5-29.4\\
 &  & 16.7-34.5
\end{array}
\right)$ \\
\hline
{\rule[-9mm]{0mm}{20mm}\bf Inverted} & $ \small \left(\begin{array}{rrr} 
48.0 & 2.8 & 3.7\\
 & 27.4 & 24.0\\
 &  & 21.7
\end{array}
\right)$ & $\small \left(\begin{array}{rrr} 
43.2-51.0 & 0.-8.6 & 0.-9.2\\
 & 21.3-31.9 & 21.3-25.6\\
 &  & 17.8-28.2
\end{array}
\right)$ & $\small \left(\begin{array}{rrr} 
 11.4-51.0 & 0.-39.0 & 0.-36.7\\
 & 0.-32.1 & 4.6-26.7\\
 &  & 0.-28.2
\end{array}
\right)$ \\
\hline
{\rule[-9mm]{0mm}{20mm}\bf Degenerate} 
 & $\small \left(\begin{array}{rrr} 
200.0 & 0.5 & 0.4\\
 & 202.7 & 2.9\\
 &  & 203.5
\end{array}\right)$ & $\small \left(\begin{array}{rrr} 
200.1-200.3 & 0.06-1.0 & 0.-1.0\\
 & 202.1-203.6 & 2.4-3.3\\
 &  & 202.5-204.1
\end{array}
\right)$ & $\small\left(\begin{array}{rrr} 
60.0-200.3 & 0.-176.6 & 0.-170.3\\
 & 0.02-203.6 & 0.5-200.3\\
 &  & 0.02-204.1
\end{array}
\right)$ \\
\hline
\end{tabular}
\caption{The reconstructed matrices for the normal and inverted mass
  hierarchies as well as  the degenerate mass spectrum for $m_0=$ 0.2 eV. 
  The moduli of elements are given
  in units of 1 meV.  For each case we present: the matrix for the
  best fit values of the mixing parameters, the intervals of the
  matrix elements obtained varying the experimental values within 1
  $\sigma$ for $\delta=\lambda_3=0$ and also for free CP-violating
  phases.  
}
\end{center}
\end{table}
}
\end{landscape}

 
\end{document}